# Separability and entanglement of n-qubits and a qubit and a qudit using Hilbert-Schmidt decompositions

Y. Ben-Aryeh * and A. Mann †

*Physics Department*,

Technion-Israel Institute of Technology, Haifa 32000, Israel

* phr65yb@physics.technion.ac.il ; † ady@physics.technion.ac.il

**Abstract**

Hilbert-Schmidt (HS) decompositions are employed for analyzing systems of n-qubits, and a qubit with a qudit. Negative eigenvalues, obtained by partial-transpose (PT) plus local unitary transformations (PTU) for one qubit from the whole system, are used for indicating entanglement/separability. A sufficient criterion for full separability of the n-qubits and qubit-qudit systems is given. We use the singular value decomposition (SVD) for improving the criterion for full separability.

General properties of entanglement and separability are analyzed for a system of a qubit and a qudit and n-qubits systems, with emphasis on maximally disordered subsystems (MDS) (i.e., density matrices for which tracing over any subsystem gives the unit density matrix). A sufficient condition that $\rho(MDS)$ is not separable is that it has an eigenvalue larger than $1/d$ for a qubit and a qudit, and larger than $1/2^{n-1}$ for n-qubits system. The PTU transformation does not change the eigenvalues of the n-qubits MDS density matrices for odd $n$. Thus the Peres-Horodecki (P-H) criterion does not give any information about entanglement of these density matrices. The P-H criterion may be useful for indicating inseparability for even n.

The changes of the entanglement and separability properties of the GHZ state, the Braid entangled state and the W state by mixing them with white noise are analyzed by the use of the present methods. The entanglement and separability properties of the GHZ-diagonal density matrices, composed of mixture of 8 GHZ density matrices with probabilities $p_i \quad (i=1,2,\cdots,8)$, is analyzed as function of these probabilities. In some cases we show that the P-H criterion is both sufficient and necessary.

## 1. Introduction

The entanglement and separability properties of 3 and 4 qubits systems have been discussed in many papers (see e.g. review articles[1,2]). In spite of the extensive literature, the use of Hilbert-Schmidt (HS) decompositions for analyzing such systems has not been largely exploited. The purpose of the present article is to analyze various interesting properties of entanglement and separability for various cases of 3 (and more) qubits systems, and combination of a qubit and a qudit, which are derived by use of such decompositions.

For 3-qubits system denoted by A, B and C we use the Hilbert-Schmidt (HS) representation of this density matrix given by:[3]

$$\rho(A,B,C)=$$
$$\frac{1}{8}\left\{\begin{array}{l}(I)_A\otimes(I)_B\otimes(I)_c+(\vec{r}\cdot\vec{\sigma})_A\otimes(I)_B\otimes(I)_c+(I)_A\otimes(\vec{s}\cdot\vec{\sigma})_B\otimes(I)_C+(I)_A\times(I)_B\otimes\left(\vec{f}\cdot\vec{\sigma}\right)_C \\ +\sum_{m,n=1}^{3}t_{mn}(I)_A\otimes(\sigma_m)_B\otimes(\sigma_n)_C+\sum_{k,l=1}^{3}o_{kl}(\sigma_k)_A\otimes(I)_B\otimes(\sigma_l)_C \\ +\sum_{i,j=1}^{3}p_{ij}(\sigma_i)_A\otimes(\sigma_j)_B\otimes(I)_C+\sum_{a,b,c=1}^{3}R_{a,b,c}(\sigma_a)_A\otimes(\sigma_b)_B\otimes(\sigma_c)_C\end{array}\right\}. \quad (1.1)$$

Here $I$ denotes the unit $2\times2$ matrix, $\otimes$ denotes outer product, the three Pauli matrices are represented by $\vec{\sigma}$; $\vec{r},\vec{s},\vec{f}$ are 3-dimensional parameters vectors, $t_{mn}, o_{kl}$, and $p_{ij}$ include 27 parameters, and 27 parameters are included for $R_{abc}$. The parameters can be obtained by tracing the product of $\rho$ by the corresponding term in (1.1). For example:

$$R_{abc}=Tr\left\{\rho(\sigma_a)_A\otimes(\sigma_b)_B\otimes(\sigma_c)_C\right\} \quad . \quad (1.2)$$

We can generalize the HS decomposition of (1.1) to that of n-qubits including $2^{2n}-1$ parameters. We concentrate in the present paper on special cases for which usually we have a smaller number of HS parameters.

For treating separability/inseparability we use the partial-transpose (PT) of a qubit.[4,5] For example, suppose the density matrix of 3-qubits $A,B$ and $C$ is fully separable, i.e.

$$\rho = \sum_j p_j \rho_A{}^{(j)} \otimes \rho_B{}^{(j)} \otimes \rho_C{}^{(j)} \qquad , \tag{1.3}$$

where $p_j \geq 0$ with $\sum_{j=1}^{N} p_j = 1$ . Then a partial transpose (say with respect to A) means replacing $\rho_A{}^{(j)}$ by $\left(\rho_A{}^{(j)}\right)^t$. Since $\rho$ and $\rho^t$ have the same eigenvalues, $\left(\rho_A{}^{(j)}\right)^t$ is a bona fide density matrix and therefore

$$\rho(PT;A) = \sum_{j=1}^{N} p_j \left(\rho_A{}^{(j)}\right)^t \otimes \rho_B{}^{(j)} \otimes \rho_C{}^{(j)} \qquad , \tag{1.4}$$

is also a density matrix, i.e., its eigenvalues are non-negative. Therefore if an eigenvalue of $\rho(PT;A)$ is negative it cannot be fully separable. [4,5]

Any $\rho_A{}^{(j)}$ can be written as

$$\rho_A{}^{(j)} = \frac{(I)_A + (\vec{a}^{(j)} \cdot (\vec{\sigma})_A)}{2} \quad ; \quad |\vec{a}^{(j)}| \leq 1 \qquad . \tag{1.5}$$

The partial transpose (PT) of $\vec{\sigma}$ $(\sigma_x, \sigma_y, \sigma_z)$ is given by $(\sigma_x, -\sigma_y, \sigma_z)$ , i.e., only $\sigma_y$ changes sign. However, a rotation (i.e. a unitary transformation which does not change the eigenvalues) by $180^0$ around the $y$ axis makes the additional transformations: $\sigma_x \rightarrow -\sigma_x$ ; $\sigma_z \rightarrow -\sigma_z$. We refer to the transformation $\vec{\sigma}_A \rightarrow -\vec{\sigma}_A$ as the PTUA transformation, i.e. partial transpose relative to A plus a local unitary transformation (of A). In a similar way we can use the PTUB or PTUC transformations relative to B or C, respectively. We should take into account that the PTU transformation can be made separately for each qubit (A, B or C), and if in any one of these cases one gets a negative eigenvalue for $\rho(PTU)$ then we can conclude that the density matrix is not fully separable. However, if in all these PTU transformations one does not get a negative eigenvalue no conclusion about entanglement can be made by this procedure. [5]

It is possible to obtain a sufficient condition for full separability of n-qubits systems by relating the HS parameters to probabilities. [6] A <u>sufficient</u> condition for full separability of the general density matrix (1.1) is given by:

$$\sum_{a,b,c=1}^{3}|R_{abc}|+\sum_{m,n=1}^{3}|t_{mn}|+\sum_{k,l=1}^{3}|o_{kl}|+\sum_{i,j=1}^{3}|p_{ij}|+|\vec{r}|+|\vec{s}|+|\vec{f}|\leq 1 \qquad , \tag{1.7}$$

i.e., if the sum of the absolute values of the HS parameters is not larger than 1, we can conclude that the density matrix is fully separable and we have an explicit expression for such separability.[6] The sufficient condition for full separability can be improved by using local unitary transformations[7,6] as the expression on the left side of (1.7) is not invariant under local unitary transformations. For a matrix $a$ the $l_p$ norm is defined as[8]

$$l_p=\left(\sum_{i=1}^{m}\sum_{j=1}^{n}|a_{ij}|^p\right)^{1/p} \quad . \tag{1.8}$$

For $l_2$ norm (1.8) gives the Frobenius norm[8] while for $l_1$ norm (1.8) gives

$$l_1=\left(\sum_{i=1}^{m}\sum_{j=1}^{n}|a_{ij}|\right) \qquad . \tag{1.9}$$

It is interesting to note that on the left side of (1.7) we have the $l_1$ norm of the matrices $t_{mn}$, $o_{kl}$, and $p_{ij}$. $\sum_{a,b,c=1}^{3}|R_{abc}|$ is considered as the $l_1$ norm of the tensor $R_{abc}$. Unitary transformations may be used to reduce the $l_1$ norm.

We analyze general properties of entanglement and separability for a system of a qubit and a qudit and n-qubits systems with maximally disordered subsystems (MDS),[7] i.e., density matrices for which tracing over any subsystem gives the unit density matrix, e.g.

$$Tr_A\rho(A,B,C)=(I)_B\otimes(I)_C/4 \quad . \tag{1.10}$$

For the 3-qubits system with MDS

$$8\rho_{ABC}=(I)_A\otimes(I)_B\otimes(I)_C+\sum_{a,b,c=1}^{3}R_{a,b,c}(\sigma_a)_A\otimes(\sigma_b)_B\otimes(\sigma_c)_C \qquad , \tag{1.11}$$

we have given an explicit form for separabity (Ref.(6) Eq.(32)) . We show that a sufficient condition that $\rho(MDS)$ is not separable is that it has an eigenvalue larger than $1/d$ for a qubit and a qudit, and larger than $1/2^{n-1}$ for n-qubits system. We prove that the PTU transformation does not change the eigenvalues of the n-qubits MDS density matrices for odd $n$. Thus the Peres-Horodecki (P-H) criterion does not give any information about entanglement of these density matrices. The P-H criterion may be useful for indicating inseparability for even n.

The present article is arranged as follows: In section 2, we discuss the use of our methods for $\rho(MDS)$[7], and also for some more general density matrices. In sections 3-6 we discuss the use of our methods for a GHZ state and for W state,[9–11] for a Braid $B_n$ entangled state,[12] and for GHZ-diagonal states,[10] including mixtures with white noise. We give sufficient conditions for separability and inseparability, and for some of these cases we show that the conditions are both sufficient and necessary. In section 7 we summarize our results and conclusions.

## 2. Mainly states with maximally disordered subsystems (MDS)

### 2.1. Some general properties of maximally disordered (MDS) qubit and qudit

The density matrix of a qubit and a qudit with maximally disordered subsystems (MDS)[7] is given in the HS decomposition by

$$2d\rho(MDS) = (I)^{2d} + \sum_{l=1}^{3}\sum_{\mu=1}^{d^2-1} t_{l,\mu}\sigma_l \otimes f_\mu \equiv (I)^{2d} + R \quad . \tag{2.1}$$

Here $\sigma_l$ are the Pauli matrices of the qubit, $(I)^{2d}$ is the unit matrix in dimension 2d and $f_\mu$ are proportional to the generators of $SU(d)$ pertaining to the qudit.[13,14] Here $f_\mu$ are $d \times d$ Hermitian traceless matrices satisfying $tr(f_\mu f_\nu) = \delta_{\mu\nu}$ . (This particular normalization is convenient for obtaining a sufficient condition for full separability). Under PTU transformation of a qubit $R \to -R$. Therefore

$$2d\rho(PTU) = (I)^{2d} - R \quad . \tag{2.2}$$

Equations (2.1) and (2.2) entail the simple relation

$$\rho\left(PTU\right)=\frac{\left(I\right)^{2d}}{d}-\rho(MDS) \quad . \tag{2.3}$$

Therefore the eigenvalues of $\rho\left(PTU\right)$ are given (in obvious notation) by

$$\lambda\left(PTU\right)=\frac{1}{d}-\lambda(\rho) \quad . \tag{2.4}$$

It follows that if eigenvalue of $\rho\,(Eq.\,2.1)$ is larger than $\frac{1}{d}$, then $\rho$ is entangled. This condition generalizes the previous condition for 2-qubits that if eigenvalue of $\rho\,(MDS)$ is larger than $\frac{1}{2}$, then $\rho$ is entangled.[7,6]

The above analysis for MDS may be generalized as follows:

Let $f_0$ be the unit matrix in dimension $d$ and suppose the density matrix is given by

$$2d\rho=\left(I\right)^{2d}+\sum_{l=1}^{3}\sum_{\mu=0}^{d^2-1}t_{l,\mu}\sigma_l\otimes f_\mu\equiv\left(I\right)^{2d}+S \quad . \tag{2.5}$$

Under PTU on the qubit

$$2d\rho\left(PTU\right)=\left(I\right)^{2d}-S \quad . \tag{2.6}$$

Therefore equations (2.3) and (2.4) hold, and if an eigenvalue of $\rho\,(Eq.\,2.5)$ is larger than $1/d$ then $\rho$ is entangled.

We now find a sufficient condition and explicit form for separability of the MDS density matrix (2.1). Performing singular value decomposition (SVD)[15,16] on the rectangular $(3\times 8)$ matrix $t_{l,\mu}$ we obtain:

$$t_{l,\mu}=\sum_{i=1}^{3}u_{li}s_iV_{i\mu} \quad . \tag{2.7}$$

Here $s_i$ are the singular values of $t_{l,\mu}$ , $u$ is $3\times3$ real orthogonal matrix and $V$ is $8\times8$ real orthogonal matrix. Hence

$$2d\rho=(I)^{2d}+\sum_{i=1}^{3}s_i\bar{\sigma}_i\otimes\bar{f}_i \quad ; \quad \bar{\sigma}_i=\sum_{l=1}^{3}u_{li}\sigma_l \quad , \quad \bar{f}_i=\sum_{\mu=1}^{8}V_{i\mu}f_\mu \quad . \tag{2.8}$$

$\bar{\sigma}$ is related to $\sigma$ by unitary transformation. It is easy to verify that $tr(\bar{f}_i\bar{f}_j)=\delta_{ij}$ . Hence the absolute values of the eigenvalues of $\bar{f}_i$ are bounded by 1. Therefore $\left[(I)^d\pm\bar{f}_i\right]/d$ is a legitimate density matrix for the qudit. We may write

$$2d\rho=\sum_{i=1}^{3}\frac{|s_i|}{2}\left[\left((I)^2+sign(s_i)\bar{\sigma}_i\right)\otimes\left((I)^d+\bar{f}_i\right)+\left((I)^2-sign(s_i)\bar{\sigma}_i\right)\otimes\left((I)^d-\bar{f}_i\right)\right]+ (I)^{2d}\left(1-\sum_{i=1}^{3}|s_i|\right) \tag{2.9}$$

and hence a sufficient condition for separability of $\rho$ is

$$\sum_{i=1}^{3}|s_i|\le1 \quad . \tag{2.10}$$

Note that $\sum_{i=1}^{3}|s_i|\le\sum_{l,m=1}^{3}|t_{lm}|$, and usually it will be a sharp inequality. The proof is similar to the one in equations (2.24-2.28) below.

### 2.2. n-qubits with maximally disordered subsystems (MDS)

We now specialize to the case that the qudit is actually composed of $n-1$ qubits, i.e. $d=2^{n-1}$ .The MDS $\rho$ is now written as

$$2^n\rho=(I)^{(n)}+\sum_{l,m,n,\cdots=1}^{3}R_{lmn\cdots}(\sigma_l)_A\otimes(\sigma_m)_B\otimes(\sigma_n)_C\cdots+\equiv(I)^{(n)}+R \quad . \tag{2.11}$$

Recall that

$$tr(\sigma_i\sigma_j)=2\delta_{ij} \quad . \tag{2.12}$$

The sufficient condition that $\rho$ is not fully separable becomes

$$\lambda(\rho) > \frac{1}{2^{n-1}} \quad . \qquad (2.13)$$

For further results about n-qubits MDS we distinguish between two cases:

**Case A:** n-qubits with maximally disordered subsystems with odd n.

By performing the (full) transpose of $\rho$ into $\rho^t$, every $\sigma_y$ in (2.11) is transformed to $-\sigma_y$. This transformation does not change the eigenvalues ($\rho$ and $\rho^t$ have the same eigenvalues). By a $180^0$ unitary rotation of all qubits around the $y$ axis the eigenvalues of the density matrix are not changed, but $\sigma_x \to -\sigma_x$ , $\sigma_y \to -\sigma_y$ . We denote the resulting density matrix by $\rho^{tu}$. Here the superscript $tu$ represents transpose of the whole density matrix, plus a unitary transformation. We emphasize that $\rho^{tu}$ and $\rho$ have the same eigenvalues. However, since we assumed an odd number of $\sigma$ we get that $R \to -R$ in this case. Hence for odd $n$

$$2^n \rho^{tu} = (I)^{(n)} - R \quad . \qquad (2.14)$$

On the other hand, the partial transpose plus a $180^0$ unitary transformation around $y$ for one qubit (say qubit A) is also given by

$$2^n \rho(PTU;A) = (I)^{(n)} - R \quad . \qquad (2.15)$$

We get the general result, that for MDS of odd $n$

$$\rho(PTU;A) = \rho^{tu} \quad . \qquad (2.16)$$

We find therefore that $\rho(PTU;A)$ has the same eigenvalues as $\rho$. Thus, for an odd number of qubits with MDS, Eq. (2.16) tells us that the P-H criterion does not give any information about entanglement.

A further conclusion comes from the fact that for an odd $n$, the eigenvalues of $(I)^{(n)} + R$ are the same as those of $(I)^{(n)} - R$; it follows that the eigenvalues of $\rho$ are $\frac{1 \pm r_i}{2^n}$ $\left(|r_i| \le 1\right)$ (The eigenvalues of $R$ come in pairs $\pm r_i$ ).

Another criterion necessary (not sufficient) for full separability of 3-qubits is[10]

$$\left|\rho_{1,8}\right| \le \left(\rho_{2,2}\rho_{3,3}\rho_{4,4}\rho_{5,5}\rho_{6,6}\rho_{7,7}\right)^{1/6} . \tag{2.17}$$

For any density matrix, semi-positivity implies

$$\left|\rho_{1,8}\right| \le \sqrt{\rho_{1,1}\rho_{8,8}} \quad . \tag{2.18}$$

For $\rho(MDS)$ we have:

$$\rho_{1,1} = \rho_{4,4} = \rho_{6,6} = \rho_{7,7} \quad ; \quad \rho_{8,8} = \rho_{2,2} = \rho_{3,3} = \rho_{5,5} . \tag{2.19}$$

Hence Eq. (2.17) is satisfied. This result may be generalized to any odd $n \quad \rho(MDS)$ .

**Case B:** n-qubits with maximally disordered subsystems with <u>even</u> n.

The argument made for case A does not apply to case B, and therefore the P-H criterion may give information on the entanglement of MDS with even n. We demonstrate it in the following simple example.

Let us assume a 4-qubits density matrix given by:

$$\begin{aligned} 16\rho = &(I)_A \otimes (I)_B \otimes (I)_C \otimes (I)_D + (\sigma_x)_A \otimes (\sigma_x)_B \otimes (\sigma_x)_C \otimes (\sigma_x)_D + \\ &(\sigma_y)_A \otimes (\sigma_y)_B \otimes (\sigma_y)_C \otimes (\sigma_y)_D + (\sigma_z)_A \otimes (\sigma_z)_B \otimes (\sigma_z)_C \otimes (\sigma_z)_D \end{aligned} \quad . \tag{2.20}$$

The 16 eigenvalues of this density matrix are

$$\lambda_i = 0 \; (i = 1, 2, \cdots, 12) \quad ; \quad \lambda_{13} = \lambda_{14} = \lambda_{15} = \lambda_{16} = 1/4 \tag{2.21}$$

Since there is an eigenvalue which is larger than $\frac{1}{2^3}$ it follows (see after Eq. (2.12) that this $\rho$ cannot be fully separable.

**2.3. Explicit full separability of 3-qubits with maximally disordered subsystems (MDS)**

For $n=3$ in (2.11) the 3-qubits products can be written (say relative to A) as

$$\sum_{l,m,n=1}^{3} R_{lmn}(\sigma_l)_A \otimes (\sigma_m)_B \otimes (\sigma_n)_C = \sum_{l=1}^{3}(\sigma_l)_A \otimes \sum_{m,n=1}^{3} R^{(l)}{}_{mn}(\sigma_m)_B \otimes (\sigma_n)_C \; ; R^{(l)}{}_{mn} \equiv R_{lmn} \; . \quad (2.22)$$

We suggest the following transformation which reduces the 27 HS parameters $R_{lmn}$ to 9 parameters with smaller $l_1$ norm. Performing the singular value decomposition (SVD)[16,17] for the matrix $R^{(l)}{}_{m,n}$ in (2.22) we get:

$$\sum_{l=1}^{3}(\sigma_l)_A \otimes \sum_{m,n=1}^{3} R^{(l)}{}_{m,n}(\sigma_m)_B \otimes (\sigma_n)_C = \sum_{l=1}^{3}(\sigma_l)_A \otimes \sum_{i=1}^{3} R^{(l)}{}_i (\sigma_i)^{(l)}{}_B \otimes (\sigma_i)^{(l)}{}_C \quad . \quad (2.23)$$

Here we used the SVD relation

$$\sum_{m,n} U^{(l)}{}_{m,i} R^{(l)}{}_{m,n} V^{(l)}{}_{j,n} = \delta_{i,j} R^{(l)}{}_i \quad (2.24)$$

where $U^{(l)}$ and $V^{(l)}$ are $3 \times 3$ real orthogonal matrices, and $(\sigma_i)_B{}^{(l)} = \sum_{m=1}^{3} U_{m,i}(\sigma_m)_B$, etc. . Taking absolute values in (2.24) we get

$$\left|R^{(l)}{}_i\right| = \left|\sum_{m,n} U^{(l)}{}_{i,m} V^{(l)}{}_{n.i} R^{(l)}{}_{m,n}\right| \le \sum_{m,n}\left|U^{(l)}{}_{i,m}\right|\left|V^{(l)}{}_{n.i}\right|\left|R^{(l)}{}_{m,n}\right| \quad (2.25)$$

Performing the summation over $i$ we get

$$\sum_{i=1}^{3}\left|R^{(l)}{}_i\right| \le \sum_{m,n=1}^{3}\sum_{i=1}^{3}\left|U^{(l)}{}_{i,m}\right|\left|V^{(l)}{}_{n.i}\right|\left|R^{(l)}{}_{m,n}\right| \quad . \quad (2.26)$$

$\left|U^{(l)}{}_{i,m}\right|$, for a certain $m$, and $\left|V^{(l)}{}_{n,i}\right|$ for a certain $n$, are unit vectors so that we get

$$\sum_{i=1}^{3}\left|U^{(l)}{}_{i,m}\right|\left|V^{(l)}{}_{n.i}\right| \le 1 \quad . \quad (2.27)$$

Substituting (2.27) into (2.26) we get the relation

$$\sum_{i=1}^{3}\left|R^{(l)}{}_{i}\right| \le \sum_{m,n=1}^{3}\left|R^{(l)}{}_{m,n}\right| \quad . \qquad (2.28)$$

We find that the sum of the singular values absolute values $\sum_{i=1}^{3}\left|R^{(l)}{}_{i}\right|$ is smaller or equal to the $l_1$ norm of the matrix $R^{(l)}{}_{m,n}$.

By substituting (2.28) into (2.22) and using the general criterion (1.7) we find that under the condition (relative to A),

$$\sum_{i,l=1}^{3}\left|R^{(l)}{}_{i}\right| \le 1 \quad , \qquad (2.29)$$

an explicit fully separable form for the 3-qubits MDS density matrix is obtained. In a similar way by using this procedure relative to B or C one gets, respectively,

$$\sum_{i,m=1}^{3}\left|R^{(m)}{}_{i}\right| \le 1 \quad ; \quad \sum_{i,n=1}^{3}\left|R^{(n)}{}_{i}\right| \le 1 \quad . \qquad (2.30)$$

One can choose the optimal condition for explicit full separability from the three conditions given by (2.29) and (2.30). A similar procedure to the above can reduce the 9 matrix elements $t_{mn}$ ($o_{kl}$ , $p_{ij}$) to 3 singular values.

**2.4. A simple example of full separability**

A very simple example of 3-qubits is

$$8\rho = (I)_A \otimes (I)_B \otimes (I)_C + R_1 (\sigma_x)_A \otimes (\sigma_x)_B \otimes (\sigma_x)_C + R_3 (I)_A \otimes (I)_B \otimes (\sigma_z)_C \quad , \qquad (2.31)$$

where here we have only two HS parameters: $R_1$ and $R_3$. In this case the eigenvalues of $8\rho$ are $1 \mp \sqrt{R_1{}^2 + R_3{}^2}$ . Therefore the condition for its being a density matrix is $R_1^2 + R_3^2 \le 1$ and in a geometric picture the eigenvalues can be represented by all points which are within the unit circle. A sufficient condition for writing it as fully-separable density matrix is: $|R_1| + |R_3| \le 1$, and In a geometrical picture this separability condition describes a square inscribed within the unit circle with vertices at points $(1,0)$ , $(0,1)$ , $(-1,0)$ , $(0,-1)$ . We should notice that the separability conditions obtained by the $l_1$ norm are sufficient but might be not necessary. In this example we can obtain full separability for all points within the circle since (2.31) can be written in the following explicitly separable form:

$$8\rho = \frac{1}{4}\begin{bmatrix} (I+\sigma_x)_A \otimes (I+\sigma_x)_B \otimes (I+R_1\sigma_x+R_3\sigma_z)_C + \\ (I+\sigma_x)_A \otimes (I-\sigma_x)_B \otimes (I-R_1\sigma_x+R_3\sigma_z)_C + \\ (I-\sigma_x)_A \otimes (I-\sigma_x)_B \otimes (I+R_1\sigma_x+R_3\sigma_z)_C + \\ (I-\sigma_x)_A \otimes (I+\sigma_x)_B \otimes (I-R_1\sigma_x+R_3\sigma_z)_C \end{bmatrix} , \qquad (2.32)$$

which holds for all cases for which $R_1^2 + R_3^2 \leq 1$.

**3. GHZ state mixed with white noise**

We would like to demonstrate by using our methods how the entanglement property of a GHZ state is changed by mixing it with a white noise.[9,17–19] For this purpose we will make the analysis for the GHZ state:

$$|\psi\rangle = \frac{1}{\sqrt{2}}\left[|0\rangle_A \otimes |0\rangle_B \otimes |0\rangle_C + |1\rangle_A \otimes |1\rangle_B \otimes |1\rangle_C\right] \quad . \qquad (3.1)$$

Similar results can be obtained for other GHZ states. We have used here the standard computational basis:

$$|0\rangle = \begin{pmatrix} 1 \\ 0 \end{pmatrix} \quad , \quad |1\rangle = \begin{pmatrix} 0 \\ 1 \end{pmatrix} \quad , \qquad (3.2)$$

as the two states of each qubit. The computational basis of 3-qubits states is described by 8-dimensional vectors where in the i'th entry we have 1 and in all other entries we have zero. Then, the density matrix 8 $\rho$ (GHZ) for the state $|\psi\rangle$ of (3.1) is given by

$$8\rho(GHZ) = 8|\psi\rangle\langle\psi| =$$

$$\begin{pmatrix} 4 & 0 & 0 & 0 & 0 & 0 & 0 & 4 \\ 0 & 0 & 0 & 0 & 0 & 0 & 0 & 0 \\ 0 & 0 & 0 & 0 & 0 & 0 & 0 & 0 \\ 0 & 0 & 0 & 0 & 0 & 0 & 0 & 0 \\ 0 & 0 & 0 & 0 & 0 & 0 & 0 & 0 \\ 0 & 0 & 0 & 0 & 0 & 0 & 0 & 0 \\ 0 & 0 & 0 & 0 & 0 & 0 & 0 & 0 \\ 4 & 0 & 0 & 0 & 0 & 0 & 0 & 4 \end{pmatrix} \quad . \tag{3.3}$$

The HS decomposition of (3.3) is given by: [3]

$$\begin{aligned} 8\rho(GHZ) = (I)_A \otimes (I)_B \otimes (I)_C + (\sigma_x)_A \otimes (\sigma_x)_B \otimes (\sigma_x)_C - (\sigma_x)_A \otimes (\sigma_y)_B \otimes (\sigma_y)_C \\ -(\sigma_y)_A \otimes (\sigma_x)_B \otimes (\sigma_y)_C - (\sigma_y)_A \otimes (\sigma_y)_B \otimes (\sigma_x)_C + (I)_A \otimes (\sigma_z)_B \otimes (\sigma_z)_C \\ +(\sigma_z)_A \otimes (I)_B \otimes (\sigma_z)_C + (\sigma_z)_A \otimes (\sigma_z)_B \otimes (I)_C \end{aligned} \quad . \tag{3.4}$$

$\rho(GHZ)$ is obviously not MDS . However, we note that any $\rho$ of n qubits may be written as

$$2^n \rho = (I)^n + G + S \quad , \tag{3.5}$$

where under a given PTU transformation (say with respect to $A$ ) $(I)^n + G$ is unchanged and $S \to -S$ . Hence

$$2^n \rho(PTU; A) = (I)^n + G - S \quad . \tag{3.6}$$

From (3.5) and (3.6) follows the general relation

$$2^n \rho(PTU; A) = 2\left[(I)^n + G\right] - 2^n \rho \quad . \tag{3.7}$$

We note that in this calculation we do not need the whole HS decomposition; we only need the $G$ part which for PTUA means those elements of the HS decomposition which include $(I)_A$ .

From Eq. (3.4) we see that $G$ in this case is given by $(I)_A \otimes (\sigma_z)_B \otimes (\sigma_z)_C$ . Then we get:

$$8\rho(GHZ;PTU;A)=2\left(I\right)^{3}+2\left(I\right)_{A}\otimes\left(\sigma_{y}\right)_{B}\otimes\left(\sigma_{y}\right)_{C}-8\rho(GHZ)=$$

$$\begin{pmatrix} 0 & 0 & 0 & 0 & 0 & 0 & 0 & -4 \\ 0 & 0 & 0 & 0 & 0 & 0 & 0 & 0 \\ 0 & 0 & 0 & 0 & 0 & 0 & 0 & 0 \\ 0 & 0 & 0 & 4 & 0 & 0 & 0 & 0 \\ 0 & 0 & 0 & 0 & 4 & 0 & 0 & 0 \\ 0 & 0 & 0 & 0 & 0 & 0 & 0 & 0 \\ 0 & 0 & 0 & 0 & 0 & 0 & 0 & 0 \\ -4 & 0 & 0 & 0 & 0 & 0 & 0 & 0 \end{pmatrix} \quad . \tag{3.8}$$

The eigenvalues of (3.8) are given by:

$$\lambda_1=\lambda_2=\lambda_3=\lambda_4=0 \quad ; \quad \lambda_5=\lambda_6=\lambda_7=4 \quad ; \quad \lambda_8=-4 \quad . \tag{3.9}$$

The negative eigenvalue $\lambda_8=-4$ indicates the inseparability property of the density matrix $\rho\left(GHZ\right)$.

Assuming that the density matrix $\rho\left(GHZ\right)$ of (3.3) with a probability $p$ is mixed with white noise with probability $(1-p)$, we get the density matrix:

$$8\rho(GHZ,mixed)=\left(1-p\right)\left(I\right)_{A}\otimes\left(I\right)_{B}\otimes\left(I\right)_{C}+p8\rho(GHZ) \quad . \tag{3.10}$$

The PTUA transformation of $8\rho(GHZ,mixed)$ is given by:

$$8\rho(GHZ,mixed,PTU;A)=\left(1-p\right)\left(I\right)_{A}\left(I\right)_{B}\left(I\right)_{C}+p8\rho(GHZ;PTU;A)=$$

$$\begin{pmatrix} 1-p & 0 & 0 & 0 & 0 & 0 & 0 & -4p \\ 0 & 1-p & 0 & 0 & 0 & 0 & 0 & 0 \\ 0 & 0 & 1-p & 0 & 0 & 0 & 0 & 0 \\ 0 & 0 & 0 & 1+3p & 0 & 0 & 0 & 0 \\ 0 & 0 & 0 & 0 & 1+3p & 0 & 0 & 0 \\ 0 & 0 & 0 & 0 & 0 & 1-p & 0 & 0 \\ 0 & 0 & 0 & 0 & 0 & 0 & 1-p & 0 \\ -4p & 0 & 0 & 0 & 0 & 0 & 0 & 1-p \end{pmatrix} \quad . \tag{3.11}$$

The eigenvalues of (3.11) are:

$$\lambda_1 = \lambda_2 = \lambda_3 = \lambda_4 = 1 - p \quad ; \quad \lambda_5 = \lambda_6 = \lambda_7 = 1 + 3p \quad ; \quad \lambda_8 = 1 - 5p \quad . \tag{3.12}$$

Under the condition $p > 1/5$ the mixed GHZ density matrix $\rho(GHZ, mixed)$ remains inseparable.

Explicit fully separable density matrices for GHZ with white noise have been derived by Schack and Caves[9] showing that for $p \le 1/5$ $\rho(GHZ, mixed)$ is fully separable. So $p \le 1/5$ is necessary and sufficient for full separability. Such result was derived also in other works[18–19] using different methods.

## 4. A Braid state mixed with white noise

The $B_n$ Braid states have been obtained[12] by the use of the $B_n$ group discovered by Artin,[20,21] where the generator operators $g_1, g_2, \cdots, g_{n-1}$ satisfy the $B_n$ Braid-group relations:

$$B_n = \left\langle \begin{matrix} g_1, g_2, \cdots, g_{n-1} \mid g_i g_j = g_j g_i & |i - j| > 1 \;; \\ g_i g_j g_i = g_j g_i g_j & |i - j| = 1 \end{matrix} \right\rangle \quad . \tag{4.1}$$

For the $B_n$ operators $g_1, g_2, \cdots, g_{n-1}$, operating on the n-qubits in our system, we use the representation:

$$\begin{aligned} g_1 &= R \otimes I \otimes I \cdots \otimes I \quad ; \\ g_2 &= I \otimes R \otimes I \cdots \otimes I \quad ; \\ g_3 &= I \otimes I \otimes R \cdots \otimes I \quad ; \\ &\cdot \\ &\cdot \\ &\cdot \\ g_{n-1} &= I \otimes I \otimes I \cdots \otimes I \otimes R \end{aligned} \quad . \tag{4.2}$$

Here $R$ is the unitary matrix given for our system by:[22]

$$R = \frac{1}{\sqrt{2}} \begin{pmatrix} 1 & 0 & 0 & 1 \\ 0 & 1 & -1 & 0 \\ 0 & 1 & 1 & 0 \\ -1 & 0 & 0 & 1 \end{pmatrix} \quad . \tag{4.3}$$

The matrix $R$ satisfies also a special Yang-Baxter equation given by:

$$(R\otimes I)\cdot(I\otimes R)\cdot(R\otimes I)=(I\otimes R)\cdot(R\otimes I)\cdot(I\otimes R) \quad . \tag{4.4}$$

Here the dot represents ordinary matrix multiplication, and $(R\otimes I)$ and $(I\otimes R)$ are matrices of $8\times 8$ dimensions. The unitary $R$ matrix can be considered in quantum computation as a universal gate related to the CNOT gate by local single qubits transformations.[22] It is quite easy to verify that the group relations (4.1) are satisfied by using (4.2) and (4.3).

The idea of using the $B_n$ operators for deriving special entangled states[12] is that by operating with the multiplication operator $g_1\cdot g_2\cdot g_3\cdots g_{n-1}$ on the computational n-qubits states $|C_i\rangle \quad (i=1,2,\cdots,2^n)$ we will get entangled states $|B_i\rangle \quad (i=1,2,\cdots,2^n)$:

$$|B_i\rangle = g_1\cdot g_2\cdots g_{n-1}|C_i\rangle \quad . \tag{4.5}$$

These states include entanglement properties analogous to those of the GHZ states, and may be considered as generalization of the Bell states[23] to large n-qubits states ($n>2$). One should take into account that $g_i \quad (i=1,2,\cdots,n-1)$ in (4.5), operate sequentially on the qubits $n,n-1\cdots,2,1$, and the matrix $R$ produces entanglement between qubits pairs $(n-1,n)\cdots(2,3),(1,2)$, sequentially.

It has been shown[12] that by operating on computational states of the 3-qubits states with the operator

$$g_1\cdot g_2=(R\otimes I)\cdot(I\otimes R) \quad , \tag{4.6}$$

we get 3-qubits Braid entangled states which have properties similar to those of the Bell 2-qubits states. In a similar way by operating on computational states of the 4-qubits with the operator

$$g_1\cdot g_2 . g_3=(R\otimes I\otimes I)\cdot(I\otimes R\otimes I)\cdot(I\otimes I\otimes R) \quad , \tag{4.7}$$

we get the 4-qubits Braid entangled states, etc. One should take into account that $g_1\cdot g_2$ is a matrix of $8\times 8$ dimension operating on 8 dimensional vectors, while $g_1\cdot g_2\cdot g_3$ is a matrix of $16\times 16$ dimension operating on 16 dimensional vectors of the computational basis of states, etc. Explicit expressions for the 3-qubits Braid entangled states have been given previously.[12]

We would like to study here how the entanglement properties of a Braid state are changed by mixing it with white noise. We make the analysis for the 3-qubits $|B1\rangle$ state. Similar results are obtained for other 3-qubis $B_n$ states. The density matrix of the $|B1\rangle$ state is given by:

$$8\rho(B1)=\begin{pmatrix} 2 & 0 & 0 & -2 & 0 & -2 & -2 & 0 \\ 0 & 0 & 0 & 0 & 0 & 0 & 0 & 0 \\ 0 & 0 & 0 & 0 & 0 & 0 & 0 & 0 \\ -2 & 0 & 0 & 2 & 0 & 2 & 2 & 0 \\ 0 & 0 & 0 & 0 & 0 & 0 & 0 & 0 \\ -2 & 0 & 0 & 2 & 0 & 2 & 2 & 0 \\ -2 & 0 & 0 & 2 & 0 & 2 & 2 & 0 \\ 0 & 0 & 0 & 0 & 0 & 0 & 0 & 0 \end{pmatrix} . \tag{4.8}$$

The HS decompositions for the density matrix $\rho(B1)$ is :

$$\begin{aligned} 8\rho(B1)) &= (I)_A (I)_B (I)_C + (\sigma_z)_A \otimes (\sigma_z)_B \otimes (\sigma_z)_C - (\sigma_x)_A \otimes (\sigma_x)_B \otimes (\sigma_z)_C \\ &- (\sigma_x)_A \otimes (\sigma_z)_B \otimes (\sigma_x)_C - (\sigma_z)_A \otimes (\sigma_x)_B \otimes (\sigma_x)_C + (I)_A \otimes (\sigma_y)_B \otimes (\sigma_y)_C \\ &+ (\sigma_y)_A \otimes (I)_B \otimes (\sigma_y)_C + (\sigma_y)_A \otimes (\sigma_y)_B \otimes (I)_C \end{aligned} . \tag{4.9}$$

Referring to Eq. (3.7), in the present case $G$ is given by $(I)_A \otimes (\sigma_y)_B \otimes (\sigma_y)_C$. Then we

$$8\rho(B1;PTU;A) = 2(I)_A \otimes (I)_B \otimes (I)_C + 2(I)_A \otimes (\sigma_y)_B \otimes (\sigma_y)_C - 8\rho(GHZ) =$$

$$\begin{pmatrix} 0 & 0 & 0 & 0 & 0 & 2 & 2 & 0 \\ 0 & 2 & 2 & 0 & 0 & 0 & 0 & 0 \\ 0 & 2 & 2 & 0 & 0 & 0 & 0 & 0 \\ 0 & 0 & 0 & 0 & 0 & -2 & -2 & 0 \\ 0 & 0 & 0 & 0 & 2 & 0 & 0 & -2 \\ 2 & 0 & 0 & -2 & 0 & 0 & 0 & 0 \\ 2 & 0 & 0 & -2 & 0 & 0 & 0 & 0 \\ 0 & 0 & 0 & 0 & 0 & 0 & 0 & 0 \end{pmatrix} \tag{4.10}$$

The eigenvalues of $8\rho(B1;PTU;A)$ are given by:

$$\lambda_1 = \lambda_2 = \lambda_3 = \lambda_4 = 0 \quad ; \quad \lambda_5 = \lambda_6 = \lambda_7 = 4 \quad ; \quad \lambda_8 = -4 \quad . \tag{4.11}$$

The negative value $\lambda_8 = -4$ indicates the entanglement of the density matrix $\rho(B1)$. Assuming that $\rho(B1)$ with a probability $p$ is mixed with white noise with probability $(1-p)$ we get the density matrix

$$8\rho(B1;mixed) = (1-p)(I)_A(I)_B(I)_C + p8\rho(B1) \quad . \tag{4.12}$$

The PTU transformation of $8\rho(B1;mixed)$ is given by

$$8\rho(B1,mixed,PTU;A) = (1-p)(I)_A(I)_B(I)_C + p8\rho(B1,PTU;A) =$$
$$\begin{pmatrix} 1-p & 0 & 0 & 0 & 0 & 2p & 2p & 0 \\ 0 & 1+p & 2p & 0 & 0 & 0 & 0 & 0 \\ 0 & 2p & 1+p & 0 & 0 & 0 & 0 & 0 \\ 0 & 0 & 0 & 1-p & 0 & -2p & -2p & 0 \\ 0 & 0 & 0 & 0 & 1+p & 0 & 0 & -2p \\ 2p & 0 & 0 & -2p & 0 & 1-p & 0 & 0 \\ 2p & 0 & 0 & -2p & 0 & 0 & 1-p & 0 \\ 0 & 0 & 0 & 0 & -2p & 0 & 0 & 1+p \end{pmatrix} . \tag{4.13}$$

The eigenvalues of $8\rho(B1,mixed,PTU;A)$ are given by:

$$\lambda_1 = \lambda_2 = \lambda_3 = \lambda_4 = 1-p \quad ; \quad \lambda_5 = \lambda_6 = \lambda_7 = 1+3p \quad ; \quad \lambda_8 = 1-5p \quad . \tag{4.14}$$

Under the condition $p > 1/5$ the Braid density matrix $\rho(B1;mixed)$ remains inseparable. For $p \le 1/5$ a fully separable form for $\rho(B1;mixed)$ is

$$\begin{aligned} &8\rho(B1;mixed) = \\ &\frac{1}{4}p \cdot \sum_{\substack{a=z,b=z,c=z \\ a=x,b=x,c=-z \\ a=x,b=-z,c=x \\ a=z,b=x,c=-x}} \left\{ \begin{matrix} (I+\sigma_a)_B \otimes (I-\sigma_b)_C \otimes (I-\sigma_c)_C + (I+\sigma_a)_A \otimes (I+\sigma_b)_B \otimes (I+\sigma_c)_C + \\ (I-\sigma_a)_A \otimes (I-\sigma_b)_B \otimes (I+\sigma_c)_A + (I-\sigma_a)_A \otimes (I+\sigma_b)_B \otimes (I-\sigma_c)_C \end{matrix} \right\} \\ &\frac{p}{2}\left[ (I+\sigma_y)_A \otimes (I+\sigma_y)_B \otimes (I+\sigma_y)_C + (I-\sigma_y)_A \otimes (I-\sigma_y)_B \otimes (I-\sigma_y)_C \right] \\ &+(1-5p)(I)_A \otimes (I)_B \otimes (I)_C \end{aligned} \quad . \tag{4.15}$$

One should notice that in deriving (4.15) we used the relation

$$p\left[(I)_A(I)_B(I)_C+t_{33}(I)_A\otimes\left(\sigma_y\right)_B\otimes\left(\sigma_y\right)_C+o_{33}\left(\sigma_y\right)_A\otimes(I)_B\otimes\left(\sigma_y\right)_C+p_{33}\left(\sigma_y\right)_A\otimes\left(\sigma_y\right)_B\otimes(I)_C\right]$$
$$=p/2\left[\left(I+\sigma_y\right)_A\otimes\left(I+\sigma_y\right)_B\otimes\left(I+\sigma_y\right)_C+\left(I-\sigma_y\right)_A\otimes\left(I-\beta\sigma_y\right)_B\otimes\left(I-\sigma_y\right)\right] \tag{4.16}$$

As in the GHZ case, $p\le 1/5$ is sufficient and necessary for full separablity. The properties of the $B_n$ entangled states are similar to those of the GHZ states as they belong to the same class.[24,25]

## 5. A W state mixed with white noise

In the present section we would like to study how the properties of the $W$ state are changed by mixing it with white noise. We will make the analysis for W state given by:

$$\left|\psi\right\rangle_W=\frac{1}{\sqrt{3}}\left[\left|0\right\rangle_A\otimes\left|0\right\rangle_B\otimes\left|1\right\rangle_C+\left|0\right\rangle_A\otimes\left|1\right\rangle_B\otimes\left|0\right\rangle_C+\left|1\right\rangle_A\otimes\left|0\right\rangle_B\otimes\left|0\right\rangle_C\right] \quad , \tag{5.1}$$

where we have used the standard computational basis. The entanglement and separability properties of the $W$ states have been treated in various works.[10,18] The explicit form for the density matrix of the $W$ state is given by:

$$3\cdot 8\rho(W)=$$
$$\begin{pmatrix} 0&0&0&0&0&0&0&0\\ 0&8&8&0&8&0&0&0\\ 0&8&8&0&8&0&0&0\\ 0&0&0&0&0&0&0&0\\ 0&8&8&0&8&0&0&0\\ 0&0&0&0&0&0&0&0\\ 0&0&0&0&0&0&0&0\\ 0&0&0&0&0&0&0&0 \end{pmatrix} \quad . \tag{5.2}$$

The HS decomposition of (5.2) is:

$$
\begin{aligned}
&3\cdot 8\rho(W)=2(\sigma_y)_A\otimes(I)_B\otimes(\sigma_y)_C+2(I)_A\otimes(\sigma_y)_B\otimes(\sigma_y)_C+2(\sigma_y)_A\otimes(\sigma_y)_B\otimes(I)_C\\
&-(\sigma_z)_A\otimes(I)_B\otimes(\sigma_z)_C-(I)_A\otimes(\sigma_z)_B\otimes(\sigma_z)_C-(\sigma_z)_A\otimes(\sigma_z)_B\otimes(I)_C\\
&+(\sigma_z)_A\otimes(I)_B\otimes(I)_C+(I)_A\otimes(\sigma_z)_B\otimes(I)_C+(I)_A\otimes(I)_B\otimes(\sigma_z)_C\\
&+2(\sigma_x)_A\otimes(\sigma_z)_B\otimes(\sigma_x)_C+2(\sigma_z)_A\otimes(\sigma_x)_B\otimes(\sigma_x)_C+2(\sigma_x)_A\otimes(\sigma_x)_B\otimes(\sigma_z)_C\\
&+2(\sigma_x)_A\otimes(\sigma_x)_B\otimes(I)_C+2(\sigma_x)_A\otimes(I)_B\otimes(\sigma_x)_C+2(I)_A\otimes(\sigma_x)_B\otimes(\sigma_x)_C\\
&+2(\sigma_y)_A\otimes(\sigma_y)_B\otimes(\sigma_z)_C+2(\sigma_y)_A\otimes(\sigma_z)_B\otimes(\sigma_y)_C+2(\sigma_z)_A\otimes(\sigma_y)_B\otimes(\sigma_y)_C\\
&-3(\sigma_z)_A\otimes(\sigma_z)_B\otimes(\sigma_z)_C+3(I)_A\otimes(I)_B\otimes(I)_C
\end{aligned} \quad . \quad (5.3)
$$

A simple way to obtain the PTA transformation of $3\cdot 8\rho(W)$ of (5.3) is to change the sign of the products in (5.3) if the first qubit is $\sigma_y$ . Then we get

$$
\begin{aligned}
&3\cdot 8\rho(W;PT;A)=3\cdot 8\rho(W)-4(\sigma_y)_A\otimes(\sigma_z)_B\otimes(\sigma_y)_C-4(\sigma_y)_A\otimes(\sigma_y)_B\otimes(\sigma_z)_C\\
&-4(\sigma_y)_A\otimes(\sigma_y)_B\otimes(I)_C-4(\sigma_y)_A\otimes(I)_B\otimes(\sigma_y)_C
\end{aligned} \quad . \quad (5.4)
$$

Using for $3\cdot 8\rho(W)$ the density matrix in the computational basis given by (5.2), and subtracting the negative terms of (5.4) we get:

$$
3\cdot 8\rho(W;PT;A)=
\begin{pmatrix}
0 & 0 & 0 & 0 & 0 & 8 & 8 & 0\\
0 & 8 & 8 & 0 & 0 & 0 & 0 & 0\\
0 & 8 & 8 & 0 & 0 & 0 & 0 & 0\\
0 & 0 & 0 & 0 & 0 & 0 & 0 & 0\\
0 & 0 & 0 & 0 & 8 & 0 & 0 & 0\\
8 & 0 & 0 & 0 & 0 & 0 & 0 & 0\\
8 & 0 & 0 & 0 & 0 & 0 & 0 & 0\\
0 & 0 & 0 & 0 & 0 & 0 & 0 & 0
\end{pmatrix} \quad . \quad (5.5)
$$

The eigenvalues of $3\cdot 8\rho(W;PT;A)$ are given by:

$$
\lambda_1=\lambda_2=\lambda_3=\lambda_4=0 \quad ; \quad \lambda_5=16 \quad ; \quad \lambda_6=8 \quad ; \quad \lambda_7=8\sqrt{2} \quad ; \quad \lambda_8=-8\sqrt{2} \quad . \quad (5.6)
$$

$\lambda_8$ indicates the inseparability of $W$ .

Assuming that the density matrix (5.2) for $\rho(W)$ with a probability $p$ is mixed with white noise with probability $(1-p)$ we get the density matrix

$$8\rho(W;mixed)=(1-p)(I)_A\otimes(I)_B\otimes(I)_C+p8\rho(W) \quad . \tag{5.7}$$

PTA transformation of $\rho(W;mixed)$ is given by:

$$3\cdot 8\rho(W;mixed;PT;A)=(1-p)(I)_A\otimes(I)_B\otimes(I)_C+p8\rho(W,PT;A)=$$

$$\begin{pmatrix} 3(1-p) & 0 & 0 & 0 & 0 & 8p & 8p & 0 \\ 0 & 3+5p & 8p & 0 & 0 & 0 & 0 & 0 \\ 0 & 8p & 3+5p & 0 & 0 & 0 & 0 & 0 \\ 0 & 0 & 0 & 3(1-p) & 0 & 0 & 0 & 0 \\ 0 & 0 & 0 & 0 & 3+5p & 0 & 0 & 0 \\ 8p & 0 & 0 & 0 & 0 & 3(1-p) & 0 & 0 \\ 8p & 0 & 0 & 0 & 0 & 0 & 3(1-p) & 0 \\ 0 & 0 & 0 & 0 & 0 & 0 & 0 & 3(1-p) \end{pmatrix} . \tag{5.8}$$

The eigenvalues of (5.8) are given by

$$\begin{aligned} &\lambda_1=\lambda_2=\lambda_3=\lambda_4=3(1-p) \;;\; \lambda_5=3+13p \;;\; \lambda_6=3+5p \;; \\ &\lambda_7=3-3p+8\sqrt{2}p \;;\; \lambda_8=3-3p-8\sqrt{2}p \end{aligned} \tag{5.9}$$

Under the condition: $p>3/(3+8\sqrt{2})\approx 0.209589$, $\lambda_8$ is negative, so that the mixed $W$ state is still inseparable. Identical condition has been derived in a different way in the literature.[18]

A fully separable form for the density matrix $8\rho(W;mixed)$ is given by:

$$\begin{aligned} &8\rho(W;mixed)=p\cdot \\ &\frac{1}{3}\begin{bmatrix} (I+\sigma_z)_A\otimes(I+\sigma_z)_B\otimes(I-\sigma_z)_C+(I-\sigma_z)_A\otimes(I+\sigma_z)_B\otimes(I+\sigma_z)_C+ \\ (I+\sigma_z)_A\otimes(I-\sigma_z)_B\otimes(I+\sigma_z)_C++(I+\sigma_x)_A\otimes(I+\sigma_x)_B\otimes(I+\sigma_z)_C+ \\ (I-\sigma_x)_A\otimes(I-\sigma_x)_B\otimes(I+\sigma_z)_C+(I+\sigma_x)_A\otimes(I+\sigma_z)_B\otimes(I+\sigma_x)_C+ \\ (I-\sigma_x)_A\otimes(I+\sigma_z)_B\otimes(I-\sigma_x)_C+(I+\sigma_z)_A\otimes(I+\sigma_x)_B\otimes(I+\sigma_x)_C+ \\ (I+\sigma_z)_A\otimes(I-\sigma_x)_B\otimes(I-\sigma_x)_C+(I+\sigma_y)_A\otimes(I+\sigma_y)_B\otimes(I+\sigma_z)_C+ \\ (I-\sigma_y)_A\otimes(I-\sigma_y)_B\otimes(I+\sigma_z)_C+(I+\sigma_y)_A\otimes(I+\sigma_z)_B\otimes(I+\sigma_y)_C+ \\ (I-\sigma_y)_A\otimes(I+\sigma_z)_B\otimes(I-\sigma_y)_C+(I+\sigma_z)_A\otimes(I+\sigma_y)_B\otimes(I+\sigma_y)_C+ \\ (I+\sigma_z)_A\otimes(I-\sigma_y)_B\otimes(I-\sigma_y)_C+4(I)_A\otimes(I)_B\otimes(I-\sigma_z)_C+ \\ 4(I)_A\otimes(I-\sigma_z)_B\otimes(I)_C+4(I-\sigma_z)_A\otimes(I)_B\otimes(I)_C \end{bmatrix} \\ &+(1-9p)(I)_A\otimes(I)_B\otimes(I)_C \end{aligned} \tag{5.10}$$

Eq. (5.10) represents $8\rho(W;mixed)$ as a fully separable density matrix in terms of products of pure states density matrices. The condition $p \le 1/9$ is therefore obviously sufficient for full separability of this density matrix. However it is not obviously necessary and perhaps may be improved by other separability methods. One should take into account that our conditions for entanglement, derived by the PTU transformations, do not distinguish between bi-separability and genuine entanglement.[10]

## 6. GHZ-diagonal states

General GHZ-diagonal states are given by

$$\rho\left(GHZ;diag\right)=\sum_{i=1}^{8} p_i \left|GHZ\right\rangle_i \left\langle GHZ\right|_i \quad . \tag{6.1}$$

Here $\left|GHZ\right\rangle_i \left\langle GHZ\right|_i \equiv (GHZ)_i$ are the 8 density matrices constructed from the 8 pure orthonormal GHZ states. $p_i$ is the probability of the corresponding density matrix. Here we treat the entanglement and separability properties of the GHZ-diagonal GHZ by the use of the HS decompositions. The properties of these states have been treated by other authors[10,11,18] using different methods. The explicit expressions for the states $(GHZ)_i$ in the computational basis are given by the following expressions, to be divided by $\sqrt{2}$ :

$$\begin{array}{l}
(GHZ)_1=\begin{pmatrix}1\\0\end{pmatrix}\otimes\begin{pmatrix}1\\0\end{pmatrix}\otimes\begin{pmatrix}1\\0\end{pmatrix}+\begin{pmatrix}0\\1\end{pmatrix}\otimes\begin{pmatrix}0\\1\end{pmatrix}\otimes\begin{pmatrix}0\\1\end{pmatrix};\,(GHZ)_2=\begin{pmatrix}1\\0\end{pmatrix}\otimes\begin{pmatrix}1\\0\end{pmatrix}\otimes\begin{pmatrix}1\\0\end{pmatrix}-\begin{pmatrix}0\\1\end{pmatrix}\otimes\begin{pmatrix}0\\1\end{pmatrix}\otimes\begin{pmatrix}0\\1\end{pmatrix}\\
(GHZ)_3=\begin{pmatrix}1\\0\end{pmatrix}\otimes\begin{pmatrix}1\\0\end{pmatrix}\otimes\begin{pmatrix}0\\1\end{pmatrix}+\begin{pmatrix}0\\1\end{pmatrix}\otimes\begin{pmatrix}0\\1\end{pmatrix}\otimes\begin{pmatrix}1\\0\end{pmatrix};\,(GHZ)_4=\begin{pmatrix}1\\0\end{pmatrix}\otimes\begin{pmatrix}1\\0\end{pmatrix}\otimes\begin{pmatrix}0\\1\end{pmatrix}-\begin{pmatrix}0\\1\end{pmatrix}\otimes\begin{pmatrix}0\\1\end{pmatrix}\otimes\begin{pmatrix}1\\0\end{pmatrix}\\
(GHZ)_5=\begin{pmatrix}1\\0\end{pmatrix}\otimes\begin{pmatrix}0\\1\end{pmatrix}\otimes\begin{pmatrix}1\\0\end{pmatrix}+\begin{pmatrix}0\\1\end{pmatrix}\otimes\begin{pmatrix}1\\0\end{pmatrix}\otimes\begin{pmatrix}0\\1\end{pmatrix};\,(GHZ)_6=\begin{pmatrix}1\\0\end{pmatrix}\otimes\begin{pmatrix}0\\1\end{pmatrix}\otimes\begin{pmatrix}1\\0\end{pmatrix}-\begin{pmatrix}0\\1\end{pmatrix}\otimes\begin{pmatrix}1\\0\end{pmatrix}\otimes\begin{pmatrix}0\\1\end{pmatrix}\\
(GHZ)_7=\begin{pmatrix}0\\1\end{pmatrix}\otimes\begin{pmatrix}0\\1\end{pmatrix}\otimes\begin{pmatrix}1\\0\end{pmatrix}+\begin{pmatrix}1\\0\end{pmatrix}\otimes\begin{pmatrix}1\\0\end{pmatrix}\otimes\begin{pmatrix}0\\1\end{pmatrix};(GHZ)_8=\begin{pmatrix}0\\1\end{pmatrix}\otimes\begin{pmatrix}0\\1\end{pmatrix}\otimes\begin{pmatrix}1\\0\end{pmatrix}-\begin{pmatrix}1\\0\end{pmatrix}\otimes\begin{pmatrix}1\\0\end{pmatrix}\otimes\begin{pmatrix}0\\1\end{pmatrix}
\end{array} \quad . \tag{6.2}$$

We assume that the probabilities $p_1, p_2, \cdots, p_8$ satisfy the relations

$$p_1 \geq p_2 \geq p_3, \cdots, \geq p_8 \quad ; \quad \sum_{i=1}^{8} p_i = 1 \quad . \tag{6.3}$$

These relations do not violate the generality of the present analysis since we can always transform the $(GHZ)_i$ $(i = 1, 2, ..., 8)$ density matrices by unitary transformations and redefine the parameters $p_i$ so that (6.3) is satisfied. The explicit expression for the density matrix (6.1) in the computational basis is

$$8\rho(GHZ; diag) = \begin{pmatrix} 4(p_1+p_2) & 0 & 0 & 0 & 0 & 0 & 0 & 4(p_1-p_2) \\ 0 & 4(p_3+p_4) & 0 & 0 & 0 & 0 & 4(p_3-p_4) & 0 \\ 0 & 0 & 4(p_5+p_6) & 0 & 0 & 4(p_5-p_6) & 0 & 0 \\ 0 & 0 & 0 & 4(p_7+p_8) & 4(p_7-p_8) & 0 & 0 & 0 \\ 0 & 0 & 0 & 4(p_7-p_8) & 4(p_7+p_8) & 0 & 0 & 0 \\ 0 & 0 & 4(p_5-p_6) & 0 & 0 & 4(p_5+p_6) & 0 & 0 \\ 0 & 4(p_3-p_4) & 0 & 0 & 0 & 0 & 4(p_3+p_4) & 0 \\ 4(p_1-p_2) & 0 & 0 & 0 & 0 & 0 & 0 & 4(p_1+p_2) \end{pmatrix} \quad (6.4$$

The HS decomposition of $8\rho(GHZ; diag)$ of (6.1,6.4), is given by:

$$\begin{gathered} 8\rho(GHZ; diag) = (I)_A (I)_B (I)_C + R_{111} (\sigma_x)_A \otimes (\sigma_x)_B \otimes (\sigma_x)_C + R_{122} (\sigma_x)_A \otimes (\sigma_y)_B \otimes (\sigma_y)_C \\ R_{212} (\sigma_y)_A \otimes (\sigma_x)_B \otimes (\sigma_y)_C + R_{221} (\sigma_y)_A \otimes (\sigma_y)_B \otimes (\sigma_x)_C + t_{33} (I)_A \otimes (\sigma_z)_B \otimes (\sigma_z)_C \\ + o_{33} (\sigma_z)_A \otimes (I)_B \otimes (\sigma_z)_C + p_{33} (\sigma_z)_A \otimes (\sigma_z)_B \otimes (I)_C \end{gathered} \quad . (6.5)$$

We have here 7 parameters: $R_{111}, R_{122}, R_{212}, R_{221}, t_{33}, o_{33}, p_{33}$.

A simple way to obtain the PTU transformation of $8\rho(GHZ; diag)$ is given by inverting the sign of the first qubit (qubit A) if it is $(\sigma_i)_A$ $(i = x, y, z)$ and leaving the sign unchanged if it is $(I)_A$ . Then the PTUA transformation of (6.5) is given as:

$$8\rho(GHZ; diag; PTU; A) = -8\rho(GHZ; diag) + 2(I)_A (I)_B (I)_C + 2t_{33} (I)_A \otimes (\sigma_z)_B \otimes (\sigma_z)_C \quad . (6.6)$$

By doing the PTU transformation relative to that of the qubit $B$ we invert the sign $(\sigma_i)_B$ $(i = x, y, z)$ but leave the sign unchanged if it is $(I)_B$ . Then we get from (6.5):

$$8\rho\left(GHZ;diag;PTU;B\right)=-8\rho\left(GHZ;diag\right)+2\left(I\right)_A\left(I\right)_B\left(I\right)_C+2o_{33}\left(\sigma_z\right)_A\otimes\left(I\right)_B\otimes\left(\sigma_z\right)_C . \quad (6.7)$$

In a similar way by doing the PTU transformation relative to that of the qubit $C$ we get:

$$8\rho\left(GHZ;diag;PTU;C\right)=-8\rho\left(GHZ;diag\right)+2\left(I\right)_A\left(I\right)_B\left(I\right)_C+2p_{33}\left(\sigma_z\right)_A\otimes\left(\sigma_z\right)_B\otimes\left(I\right)_C . \quad (6.8)$$

By substituting $8\rho\left(GHZ;diag\right)$ from (6.4) into (6.6) we get

$$8\rho\left(GHZ;diag;PTUA\right)=$$

$$\begin{pmatrix}
\begin{matrix}-4(p_1+p_2)\\2t_{33}+2\end{matrix} & 0 & 0 & 0 & 0 & 0 & 0 & -4(p_1-p_2) \\
0 & \begin{matrix}-4(p_3+p_4)\\-2t_{33}+2\end{matrix} & 0 & 0 & 0 & 0 & -4(p_3-p_4) & 0 \\
0 & 0 & \begin{matrix}-4(p_5+p_6)\\2t_{33}+2\end{matrix} & 0 & 0 & -4(p_5-p_6) & 0 & 0 \\
0 & 0 & 0 & \begin{matrix}-4(p_7+p_8)\\-2t_{33}+2\end{matrix} & -4(p_7-p_8) & 0 & 0 & 0 \\
0 & 0 & 0 & -4(p_7-p_8) & \begin{matrix}-4(p_7+p_8)\\-2t_{33}+2\end{matrix} & 0 & 0 & 0 \\
0 & 0 & -4(p_5-p_6) & 0 & 0 & \begin{matrix}-4(p_5+p_6)\\2t_{33}+2\end{matrix} & 0 & 0 \\
0 & -4(p_3-p_4) & 0 & 0 & 0 & 0 & \begin{matrix}-4(p_3+p_4)\\-2t_{33}+2\end{matrix} & 0 \\
-4(p_1-p_2) & 0 & 0 & 0 & 0 & 0 & 0 & \begin{matrix}-4(p_1+p_2)\\2t_{33}+2\end{matrix}
\end{pmatrix}$$

(6.9)

By evaluating all the products of (6.5) in the computational basis and comparing them with the density matrix (6.4) we get after straightforward calculations the relations:

$$R_{111}=p_1+p_3+p_5+p_7-p_2-p_4-p_6-p_8\ ;\ R_{221}=p_2+p_4+p_5+p_7-p_1-p_3-p_6-p_8\ ;$$
$$R_{212}=p_2+p_3+p_6+p_7-p_1-p_4-p_5-p_8\ ;\ R_{122}=p_2+p_3+p_5+p_8-p_1-p_4-p_6-p_7\ ; \quad (6.10)$$
$$t_{33}+1=2\left(p_1+p_2+p_7+p_8\right);\ 0_{33}+1=2\left(p_1+p_2+p_5+p_6\right);$$
$$p_{33}+1=2\left(p_1+p_2+p_3+p_4\right)$$

Eigenvalues of (6.9) are obtained by solving 4 quadratic equations. By taking into account in (6.9) the terms in the first and eight lines with the corresponding terms in the first and eight columns we get the eigenvalues for the quadratic equation

$$\left[-4(p_1+p_2+2+2t_{33}-\lambda)\right]^2=4\left(p_1-p_2\right)^2 \quad . \tag{6.11}$$

Using the relation $2t_{33}+2=4\left(p_1+p_2+p_7+p_8\right)$ obtained from (6.10), and due to the use of the relation (6.3) we find that the smallest eigenvalue of (6.11) is obtained from this equation and is given by:

$$\lambda_{\min}(PTU;A)=4\left[p_7+p_8-\left(p_1-p_2\right)\right] \quad . \tag{6.12}$$

So that a negative value of $\lambda_{\min}(PTUA)$ , i.e.

$$p_7+p_8-\left(p_1-p_2\right)<0 \qquad , \tag{6.13}$$

is a sufficient condition for the density matrix $8\rho\left(GHZ;diag\right)$ to be inseparable, with respect to A.

By substituting $8\rho\left(GHZ;diag\right)$ from (6.4) into (6.7) and transforming this equation into the computational basis we obtain equations which are similar to (6.9), (6.10) and (6.11) and get the minimal value for the PTUB transformation:

$$\lambda_{\min}(PTU;B)=4\left[p_5+p_6-\left(p_1-p_2\right)\right] \quad . \tag{6.14}$$

So that a negative value of $\lambda_{\min}(PTU;B)$, i.e.

$$p_5+p_6-\left(p_1-p_2\right)<0 \quad , \tag{6.15}$$

is a sufficient condition for the density matrix $8\rho\left(GHZ;diag\right)$ to be inseparable with respect to B.

By substituting $8\rho\left(GHZ;diag\right)$ from (6.4) into (6.8) and transforming this equation into the computational we get in a similar way the relation

$$\lambda_{\min}(PTU;C)=4\left[p_3+p_4-\left(p_1-p_2\right)\right]. \tag{6.16}$$

So that a negative value for $\lambda_{\min}(PTU;C)$ , i. e. ,

$$p_3+p_4-\left(p_1-p_2\right)<0 \quad , \tag{6.17}$$

Is a sufficient condition for the density matrix $8\rho\left(GHZ;diag\right)$ to be inseparable with respect to C.

The condition (6.17) for entanglement implies also the conditions (6.15) and the condition (6.13), but not vice versa.

A sufficient condition for separability can be given by applying the general equation (1.7) to the GHZ-diagonal states obtaining

$$\left|R_{111}\right|+\left|R_{122}\right|+\left|R_{212}\right|+\left|R_{221}\right|+\left|t_{33}\right|+\left|o_{33}\right|+\left|p_{33}\right|\leq 1 \quad . \tag{6.18}$$

We have here 7 parameters. We would like to apply the relations (6.10) in (6.18), but in order to do this we have to clarify what are the absolute values of the parameters given by (6.10). From (6.3) it is straightforward to notice that the parameters $o_{33}, p_{33}$ and $R_{111}$ are non-negative i.e.

$$o_{33}\geq 0 \quad , \quad p_{33}\geq 0 \quad , \quad R_{111}\geq 0 \quad . \tag{6.19}$$

For finding separability conditions for $8\rho\left(GHZ;diag\right)$ we assume that the condition (6.17) for entanglement is not satisfied, so that we can assume for treating separability

$$\left(p_1-p_2\right)\geq p_3+p_4 \quad . \tag{6.20}$$

By using Eq. (6.3) and substituting the relations (6.20) into the equations for $R_{122}$ , $R_{212}$ , $R_{221}$ and $t_{33}$, we find after some simple algebra that the parameters $R_{122}$ , $R_{212}$ , $R_{221}$ are non-positive while $t_{33}$ is positive , i.e.,

$$t_{33}\geq 0 \quad , \quad R_{122}\leq 0 \quad , \quad R_{212}\leq 0 \quad , \quad R_{221}\leq 0 \quad . \tag{6.21}$$

Using equations (6.18), (6.19) and (6.21) we obtain a sufficient condition for the separability of the GHZ –diagonal states.

$$R_{111}-R_{122}-R_{212}-R_{221}+t_{33}+o_{33}+p_{33}\leq 1 \quad . \tag{6.22}$$

By substituting in (6.22) the present 7 HS parameters from (6.10) and using (6.3) we get after straightforward calculations,

$$\begin{gathered} 10p_1+2p_2+2p_3+2p_4+2p_5+2p_6+2p_7+2p_8-3\leq 1 \quad ; \\ 8p_1\leq 2\rightarrow p_1\leq 1/4 \end{gathered} \tag{6.23}$$

The condition $p_1 \le 1/4$ is therefore obviously a sufficient condition for full separability of the GHZ-diagonal density matrices. However it is not necessary and may be improved by other separability methods.[10,18] In particular let us assume the special case

$$t_{33} = o_{33} = p_{33} = C \quad . \tag{6.24}$$

Then from (6.3) and (6.10) we get:

$$\begin{aligned} & p_3 = p_4 = p_5 = p_6 = p_7 = p_8 = \frac{1-p_1-p_2}{6} \quad , \\ & C = \frac{4(p_1+p_2)-1}{3} \ge 0 \quad , \quad . \\ & R_{212} = R_{221} = R_{122} = -R_{111} = p_2 - p_1 \end{aligned} \tag{6.25}$$

Using (6.24) we have

$$\begin{aligned} & \left[ t_{33}(I)_A \otimes (\sigma_y)_B \otimes (\sigma_y)_C + o_{33}(\sigma_y)_A \otimes (I)_B \otimes (\sigma_y)_C + p_{33}(\sigma_y)_A \otimes (\sigma_y)_B \otimes (I)_C \right] = \\ & C/2\left[ (I+\sigma_y)_A \otimes (I+\sigma_y)_B \otimes (I+\sigma_y)_C + (I-\sigma_y)_A \otimes (I-\beta\sigma_y)_B \otimes (I-\sigma_y) \right] - CI \otimes I \otimes I \end{aligned} \tag{6.26}$$

The criterion for full separability (6.18) becomes

$$|R_{111}| + |R_{122}| + |R_{212}| + |R_{221}| + C \le 1 \quad . \tag{6.27}$$

Substituting (6.25) in (6.27) we get a sufficient condition for full separability

$$p_1 - p_2/2 \le 1/4 \quad . \tag{6.28}$$

The condition for inseparability (6.17) for this case becomes

$$\left[ \frac{1-p_1-p_2}{3} < p_1 - p_2 \right] \to p_1 - p_2/2 > 1/4 \quad . \tag{6.29}$$

Therefore the P-H criterion in this case is both necessary and sufficient for inseparability, when (6.24) is satisfied. This result is in agreement with [17}.

We notice that in the present case the condition for full separability is improved relative to (6.23). In particular, for the case of mixture of GHZ with white noise (section 3):( $p_1 = \frac{1+7p}{8}$ , $p_2 = p_3 = \cdots = p_8 = \frac{1-p}{8}$ ) Eq.(6.28) yields $p \le 1/5$ in agreement with the analysis of section (3).

**7. Summary**

In the present work separability/entanglement properties of multiple qubits systems and a qubit and a qudit were treated by the use of HS decompositions. It is straightforward to do the PT transformation which amounts to changing the sign of $\sigma_y$ , of one qubit, or do the PTU transformation which amounts to changing the sign of $\sigma_i\ (i = x, y, z)$ for one qubit. If for any density matrix we get by PT or PTU transformations negative eigenvalues we can conclude that the original density matrix is not fully separable.

In section 2 we treated the density matrices with MDS[7] given in the HS decompositions by (2.1) and (2.11). For MDS n-qubits system with odd $n$ the PTU transformation does not change the eigenvalues, so that the P-H criterion for these density matrices is mute. For even n-qubits systems, the PTU transformation can give negative eigenvalues and thus can give information on the inseparability of the n-qubits systems. We have demonstrated for a simple example given by (2.17) that the use of the Peres criterion shows that it is inseparable. We analyzed the entanglement and separability properties of a MDS system composed of a qubit and a qudit. By using SVD we find that in analogous way to that of the 2-qubits, we get three parameters $s_1, s_2, s_3$ where a sufficient condition for separability is given by (2.10). We find also that if an eigenvalue of the MDS density matrix $\rho_{AB}$, composed of a qubit plus qudit, is larger than $1/d$ , then $\rho_{AB}$ is entangled.

We used the present method to analyze the entanglement properties of GHZ state, Braid $B_n$ entangled state[12] and $W$ state mixed with white noise given by $\rho(mixed) = p\rho + (1-p)\left[(I)_A \otimes (I)_B \otimes (I)_C / 8\right]$, where $p$ is the probability for the original density matrix $\rho$ , and $(1-p)$ is the probability for the white noise. We calculated the minimal value of $p$ , for which the PTU transformation still yields negative eigenvalue, hence indicating inseparability. We found that under the condition $p > 1/5$ the mixed GHZ and Braid state remain entangled as derived in (3.12) and (4.14), respectively, while for $p \le 1/5$ they are fully separable. (The Peres-Horodecki criterion is necessary and sufficient). Under the condition $p > \frac{3}{3+8\sqrt{2}} \approx 0.20958$ the mixed $W$ state remains entangled as derived in (5.9).

The HS decomposition derived in (5.3) for the $W$ density matrix has been transformed in (5.10) to a quite complicated fully separable density matrix, composed of products of single qubits pure states density matrices. We found that the condition $p \leq 1/9$ is obviously sufficient for full separability but perhaps may be improved by other separability methods.[17–18]

In section 6 we used our methods for treating entanglement and separability properties of GHZ-diagonal states. We have taken into account that the GHZ-diagonal density matrices are not symmetric relative to the exchange between the qubits A, B and C. Then, by performing the PTU transformations for the qubits A, B and C we obtained the corresponding sufficient conditions for inseparability given by (6.13), (6.15) and (6.17), respectively. We find that a sufficient condition for separability of the GHZ diagonal states is given by the simple relation $p_1 \leq 1/4$. This condition is obviously a sufficient condition for full separability but might be improved by other separability methods.[17,18] For the case of two independent parameters $p_1$ and $p_2$, and $p_3 = p_4 = \cdots = p_8$ ,the criterion for inseparability $p_1 - p_2/2 > 1/4$ is both necessary and sufficient.

**References**

1. R. Horodecki, P. Horodecki, M. Horodecki, and K. Horodecki, Quantum Entanglement, *Rev. Mod. Phys.* **81** (2009) 865-942.
2. O. Guhne and G. Toth, Entanglement Detection, *Physics Reports* **474** (2009) 1-75.
3. Y. Ben-Aryeh, A. Mann and B.C. Sanders, Empirical state determination of entangled two-level systems and its relation to information theory, *Foundations of Physics* **29** (1999) 1963-1975.
4. A. Peres, Separability criterion for density matrices., *Phys. Rev. Lett.* **77** (1996) 1413-1415.
5. M. Horodecki , P. Horodecki and R. Horodecki, Separability of mixed states: necessary and sufficient conditions, *Phys. Lett.* A **223** (1996) 1-8.
6. Y. Ben-Aryeh and A. Mann, Explicit constructions of all separable two-qubits density matrices and related problems for three-qubits systems, *International Journal of Quantum Information* **13** (2015) 1550061; arXiv.org: 1510.07222
7. R. Horodecki and M. Horodecki, Information-theoretic aspects of inseparability of mixed states, *Phys. Rev.* A **54** (1996) 1838-1843.
8. R. Horn and C. R. Johnson, *Matrix Analysis,* (Cambridge University Press, Cambridge, 1991).
9. R. Schack and and C. M. Caves , Explicit product ensembles for separable quantum states, *J. Mod. Opt.* **47** (2000) 387-398; arXiv.org: quant-ph/9904109v2 May (1999).

10. O. Guhne and M. Seevinck, Separability criteria for genuine multi-particle entanglement, *New Journal of Physics* **12** (2010) 053002 .
11. O. Guhne, Entanglement criteria and full separability of multi-qubit quantum states, *Phys. Lett. A* **375**, (2011). 406-410 .
12. Y. Ben-Aryeh , Entangled states implemented by $B_n$ group operators, including properties based on HS decompositions , separability and concurrence , *International Journal of Quantum Inf.* **13** (2015) 1450045 ; The use of Braid operators for implementing entangled large n-qubits Bell states, arXiv.org (quant- ph) 1403.2524.
13. B. C. Hall, *Lie Groups, Lie Algebras, and Representations: An Elementary Introduction* (Springer, New York, 2003).
14. A. Zee, *Group Theory in a Nutshell fot Physicists* (Princeton University Press , Princeton 2016).
15. L. De Lathauwer, B. De Moor, and J. Vanderwalle, A multilinear singular value decompossition, SIAM J. Matrix Anal. Appl. **21** (2000) 1253-1278.
16. G. H. Golub, C. F. Van Loan, *Matrix Computation, Fourth Edition* (John Hopkins University Press, Baltimore, 2013).
17. L E. Buchholtz, T. Moroder, and O. Guhne, Evaluating the geometric measure of multiparticle entanglement, Ann. Phys. (Berlin) **528** (2016) 278-287 .
18. S. Szalay, Separability criteria for mixed three-qubits states, *Phys. Rev.* A **83** (2011) 062337.
19. W. Dür and J. I. Cirac, Classification of multiqubit mixed states : Separability and distillability properties, *Phys. Rev.* A **61** (2000) 042314.
20. E. Artin. Theory of Braids, *Annals of Mathematics* **48** (1947) 101-126.
21. J. Birman . Braids, Links and Mapping Class Groups, *Annals of Mathematics Studies, No.* 82 (Princeton University Press, Princeton , 1975).
22. L. H. Kauffman and S. J. Jr. Lomonaco, Braiding operators are universal quantum gates, *New Journal of Physics* **6** (2004) 134:1-40.
23. S .L. Braunstein, A. Mann and M. Revzen, Maximal violation of Bell inequalities for mixed states, *Phys. Rev. Lett.* (1992) 3259-3261.
24. A. Acin, D. BruB, M. Lewenstein and A. Sanpera, Classification of mixed three-qubits states, *Phys. Rev. Lett.* **87** (2001) 040401.
25. W. Dür , G. Vidal and J.I. Cirac, Three qubits can be entangled in two inequivalent ways, *Phys. Rev. A* **62** (2000) 062314.